\documentclass[preprint,journal]{vgtc}       

\ifpdf
  \pdfoutput=1\relax                   
  \usepackage{graphicx}                
  \DeclareGraphicsExtensions{.pdf,.png,.jpg,.jpeg} 
\else
  \usepackage{graphicx}                
  \DeclareGraphicsExtensions{.eps}     
\fi%

\graphicspath{{figures/}{pictures/}{images/}{./}} 

\PassOptionsToPackage{warn}{textcomp}  
\usepackage{textcomp}                  
\usepackage{times}                     
\usepackage{cite}                      
\usepackage{tabu}                      
\usepackage{booktabs}                  
\usepackage{xspace}

\usepackage{paralist}
\usepackage{enumitem}
\usepackage{gensymb}
\usepackage{amsmath}
\usepackage{amssymb}

\usepackage{moresize}

\usepackage{wrapfig}
\usepackage{subcaption}

\usepackage{listings}
\newlength{\listingindent}                
\lstdefinelanguage{goh}
{morekeywords={in,&,<,>,=},
sensitive=false,
morecomment=[l]{//},
morecomment=[s]{/*}{*/},
morestring=[b]",
}

\onlineid{1203}

\vgtccategory{Research}
\vgtcpapertype{system}

\usepackage[dvipsnames]{xcolor}
\usepackage[normalem]{ulem}
\definecolor{mred}{rgb}{.80,.12,.30}
\definecolor{MRED}{rgb}{.80,.12,.30}
\definecolor{grey}{rgb}{0.5,0.5,0.5}
\definecolor{lgrey}{rgb}{0.7,0.7,0.7}
\definecolor{purple}{rgb}{.75,0,.85}
\definecolor{pistachio}{rgb}{0.58, 0.77, 0.45}
\definecolor{myorange}{rgb}{0.94, 0.36, 0.13}

\newif\ifnotes
\notesfalse

\newcommand{\mib}[1]{\ifnotes{\textcolor{myorange}{(Michael: #1)}}\else{#1}\fi}

\let\origcite\cite
\renewcommand{\cite}[1]{\ifnotes\mbox{\origcite{#1}}\else \origcite{#1}\fi}

\usepackage[activate={true,nocompatibility},final,tracking=true,kerning=true,spacing=true,factor=900,stretch=25,shrink=10]{microtype}

\newcommand{\sys}{{PIXAL}\xspace}
\newcommand{\alg}{{RPI algorithm}\xspace}
\newcommand{\sysvis}{{PixalExplorer}\xspace}
\newcommand{\sysreport}{{Pixalate}\xspace}

\title{PIXAL: Anomaly Reasoning with Visual Analytics}

\author{Brian Montambault, Camelia D. Brumar, Michael Behrisch, Remco Chang}
\authorfooter{
\item
Brian Montambault is with Tufts University. E-mail: brian.montambault@tufts.edu.
}

\shortauthortitle{Montambault \MakeLowercase{\textit{et al.}}: \alg}

\abstract{
Anomaly detection remains an open challenge in many application areas. While there are a number of available machine learning algorithms for detecting anomalies, analysts are frequently asked to take additional steps in reasoning about the root cause of the anomalies and form actionable hypotheses that can be communicated to business stakeholders. Without the appropriate tools, this reasoning process is time-consuming, tedious, and potentially error-prone. In this paper we present PIXAL, a visual analytics system developed following an iterative design process with professional analysts responsible for anomaly detection. PIXAL is designed to fill gaps in existing tools commonly used by analysts to reason with and make sense of anomalies. PIXAL consists of three components: (1) an algorithm that finds patterns by aggregating multiple anomalous data points using first-order predicates, (2) a visualization tool that allows the analyst to build trust in the algorithmically-generated predicates by performing comparative and counterfactual analyses, and (3) a visualization tool that helps the analyst generate and validate hypotheses by exploring which features in the data most explain the anomalies. Finally, we present the results of a qualitative observational study with professional analysts. These results of the study indicate that PIXAL facilitates the anomaly reasoning process, allowing analysts to make sense of anomalies and generate hypotheses that are meaningful and actionable to business stakeholders.

} 


\CCScatlist{ 
 \CCScat{K.6.1}{Management of Computing and Information Systems}%
{Project and People Management}{Life Cycle};
 \CCScat{K.7.m}{The Computing Profession}{Miscellaneous}{Ethics}
}

\teaser{
 \centering
 \includegraphics[width=\linewidth]{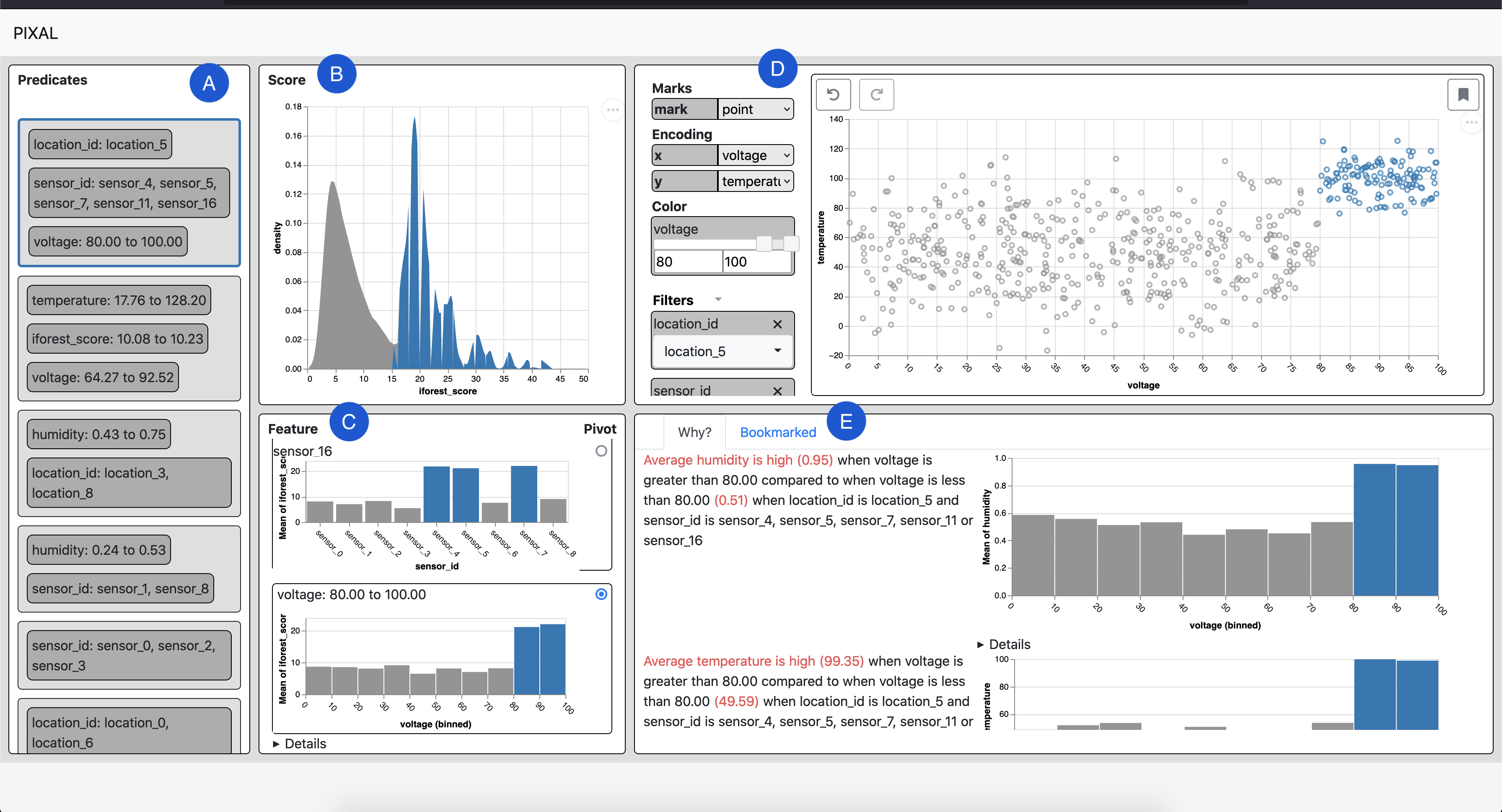}
 \caption{Interface of \sysreport, one of the two visualization tools in \sys. \sysreport helps an analyst prepare their report to stakeholders and decision makers. It consists of five views: (A) Predicate View that shows all predicates; (B) Distribution View shows the distributions of anomaly scores for the selected predicate and its complement; (C) Pivot View allows the analyst to pivot on one of the features in a predicate; (D) Exploration View helps an analyst explore different visualizations of a predicate; and (E) Recommendation View shows possible explanations generated by the system that might explain the anomalies. 
 }
	\label{fig:systemoverview}
}

\vgtcinsertpkg

\begin{document}

\abovedisplayskip=6pt
\abovedisplayshortskip=0pt
\belowdisplayskip=6pt
\belowdisplayshortskip=6pt


\firstsection{Introduction}
\maketitle

Anomaly detection is a critical part of many domains today, with organizations attempting to identify potential threats, failures, or other significant deviations from normal behavior.
A number of machine learning algorithms for detecting anomalies are available (e.g., see\cite{Chandola2009ADSurvey, chalapathy2019deep}).
These algorithms are effective in assessing and assigning every row in a dataset an \textit{anomaly score} such that data points with scores above a threshold are deemed anomalous.
However, knowing which data points are anomalous is not enough -- business stakeholders and decision makers need higher-level reasoning to understand the patterns and potential root causes of the anomalies before they can take meaningful actions.

Unfortunately, few tools exist to support the task of \textit{reasoning} or making sense of anomalous data. \mib{call it anomaly reasoning (fig2) and state why important}
Instead, data analysts currently need to manually inspect the anomalous data points, create visualizations to discover potential patterns, mentally construct plausible hypotheses to describe the patterns, and generate reports in ways that are understandable and meaningful to the business stakeholders and decision makers.
Without the appropriate support, these tasks can be tedious, time-consuming, and potentially error-prone. 

\begin{figure*}[t!]
 \centering
 \includegraphics[width=0.9\linewidth]{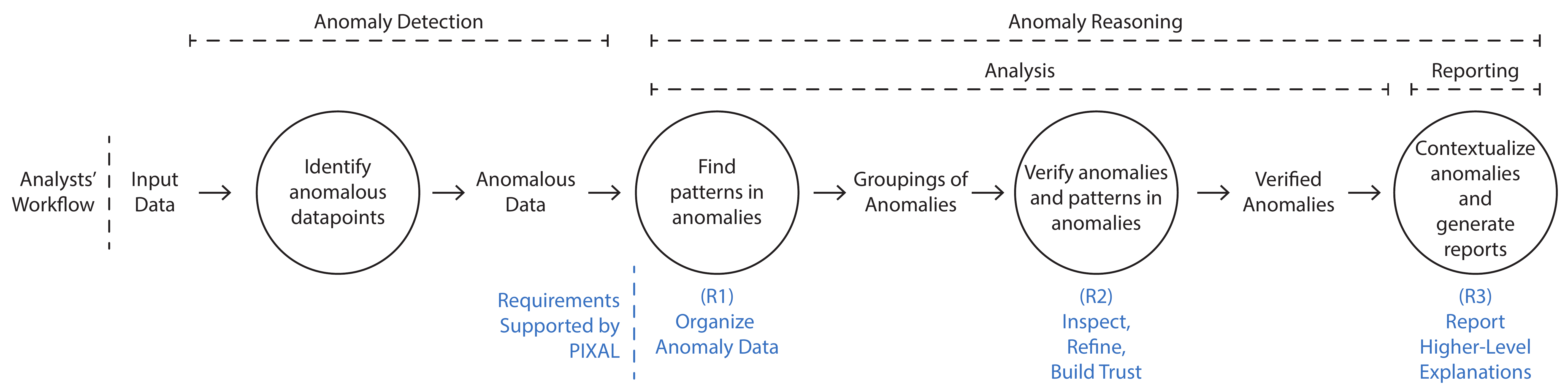}
\caption{The anomaly reasoning workflow as informed by data analysts, which consists of two stages of work: Anomaly Detection and Anomaly Reasoning (see Section~ \ref{sec:motivation}). First, the analyst identifies individual anomalies in the data.
Next, they organize anomalies into groups, test and verify the validity of those groups, and finally propose a hypothesis and report their findings to stakeholders. While a number of accessible tools exist for anomaly detection, analyst report that the full series of tasks required for anomaly \textit{reasoning} is not well supported by existing tools. We propose \sys as an integrated system supporting the three stages of anomaly reasoning shown above.}
	\label{fig:analystWorkflowAndRequirements}
\end{figure*}

To address this unmet need, in this paper\mib{can be deleted ``in this paper''}, we present \sys, a visual analytics system designed to help an analyst with the task of \textit{anomaly reasoning}.
\sys is developed following an iterative design process with professional data analysts whose jobs involve anomaly detection, reasoning, and reporting. 
Based on the interviews and the feedback from the analysts, \sys is designed with three requirements in mind.
First, an anomaly reasoning tool must help an analyst organize and make sense of the collection of anomalous data points. 
Second, the anomaly reasoning tool must assist the analyst in building trust in the found anomalies.
Lastly, the anomaly reasoning tool must support the analyst in reporting their findings to the business stakeholders and decision makers such that they can take meaningful actions.

To fulfill these design requirements, \sys consists of three components.
The first is the recursive predicate induction (RPI) algorithm for discovering groupings of anomalous data points.
Similar to a regression tree\cite{lewis2000cart}, the \alg takes data points and their assigned anomaly scores and outputs groupings of these points.
In particular, the \alg expresses these groupings as first-order predicates (e.g., \texttt{(city='Boston') \& (precipitation > 7.6) \& (month in [Nov, Dec, Jan])}), and has two unique properties that can aid anomaly reasoning. %
First, the \alg can return \textit{multiple} groups of anomalies (as predicates) that may overlap the same data points. 
Having access to different groupings can help analysts reason about how anomalies might be related to multiple different underlying phenomena. 
Second, the \alg provides the likelihood that data points in grouping have high anomaly scores.
Comparing the data points in a grouping to the rest of the data with a Bayesian t-test\cite{BayesTtest} produces a Bayes factor corresponding to the degree of evidence that each predicate contains anomalies.

The second component of \sys is the \sysvis tool that allows analysts to fine-tune and build trust in the predicates generated by the \alg.
At its core, \sysvis is an engine that can visualize any first-order predicate as a histogram that shows the distribution of anomaly scores for the data points it contains.
Analysts can make modifications to a predicate produced by the \alg to perform counterfactual analyses, superimpose two or more predicates to make comparisons, 
or refine the predicate to better reflect their domain knowledge.
\sysvis provides an interface that supports the analyst in specifying and managing the predicates, and GUI elements (such as sliders) to reduce the burden of repetitive typing during the analysis. 

Finally, the third component of \sys is \sysreport, a visualization tool to help analysts generate hypotheses about observed anomalies that can be used in a report for business stakeholders and decision makers.
\sysreport assists the analyst in making sense of anomaly scores.
Rather than reporting predicates directly, which can describe similar data points with high anomaly scores but not why their anomaly scores are high, analysts can use \sysreport
to produce reports with higher-level semantic meaning. 
For example, for the predicate \texttt{(city='Boston') \& (precipitation > 7.6) \& (month in [Nov, Dec, Jan])}, using \sysreport a business analyst might find that the anomaly highly correlates with a decrease in sales and deduce that the anomalies are likely due to slow sales during winter storms in Boston.



To evaluate the effectiveness of \sys in supporting analysts' anomaly reasoning,
we conducted an interview study with three professional data analysts whose jobs involve anomaly detection.
The analysts reported 
that they found \sys to be effective in facilitating reasoning and hypothesis-generation, suggesting the potential of \sys as a tool for helping them perform anomaly reasoning with real-world problems.
In summary, our work on \sys makes the following contributions:

\begin{itemize}[topsep=6pt, partopsep=0pt,itemsep=0pt,parsep=0pt]
\item We introduce \sys, a visual analytics system that supports anomaly reasoning.
\item \sys consists of three integrated components that together  help analysts aggregate and refine anomalous data points into higher-level, semantically meaningful reports. 
\item We conducted an interview study with three professional data analysts who validated the effectiveness of \sys.
\end{itemize}

\section{Related Work}
We describe prior research in explainable anomaly detection and visual analytics related to our work.


\subsection{Anomaly Detection and Explanation}

Anomaly detection algorithms typically output \textit{anomaly scores} that indicate how anomalous each instance is, then classify any instance with an anomaly score above a set threshold as an anomaly.
Although the specific methods differ, all of these algorithms aim to calculate the degree to which each data point deviates from the rest of the data.
For example, the local outlier factor algorithm\cite{breunig2000lof} calculates the average reachability of each point relative to its $k$ nearest neighbors, while the isolation forest (iforest) algorithm\cite{liu2008isolation} calculates the number of splits needed by a random forest to isolate each data point. For a comprehensive review of anomaly detection techniques in machine learning and deep learning, see surveys by Chalapathy et al.\cite{Chandola2009ADSurvey} and Vandola et al.\cite{chalapathy2019deep}.

In addition to anomaly detection, a number of approaches have been proposed in the Machine Learning (ML) and the Explainable Artificial Intelligence (XAI) communities for explainable anomaly detection.
Lipton suggests two classes of explanations: \textit{local} explanations of a model's decision on individual data instances and \textit{global} explanations of the models general behavior\cite{lipton2018mythos}.
Examples of local-explanation approaches include work by Amarasinghe et al.\cite{Amarasinghe2018DNNLocalXAD}, who use a deep neural network for detecting anomalies and generating feature relevance scores for each instance.
Marino et al.\cite{Marino2018AdversarialLocalXAD} propose an approach that can provide local explanations for any anomaly detection model by reporting the minimum modification of a misclassified instance's features that would cause it to be correctly classified.
Antwarg et al.\cite{Antwarg2019SHAPLocalXAD} generate local explanations for an autoencoder-based anomaly detection model by using SHAP (SHapley Additive exPlanations) values to compute the contribution of each feature to the models decision on a particular instance.
LIME\cite{DBLP:conf/kdd/Ribeiro0G16} generates local explanations for individual predictions by approximating a model's decision function with an explainable model trained on a local subspaces.

The advantage of local-explanation approaches is that it can help an analyst detect and diagnose a failure point -- for example, identifying a erroneous feature in an ML model or a bad actor in social-network analysis. 
However, conversely local-explanation approaches are less effective in helping analysts with higher-level reasoning. 
In cases where the analyst needs to understand a phenomenon or a reason behind the anomalies, global-explanation approaches may be more appropriate.
%
%
For example, TrafficAV\cite{Wang2016TrafficGlobalXAD} combines network traffic analysis with a decision tree model to detect malware behavior in network traffic generated by mobile applications.
This system generates feature weights that explain the characteristics of each family of malware.
A similar approach is taken by Carletti et al.\cite{Carletti2019FeatureImportanceGlobalXAD} which generates weights to explain which features contribute to anomaly scores assigned by the commonly used isolation forest algorithm.
Nguyen et al.\cite{Nguyen2019GradientFeatureImportanceGlobalXAD} propose a method for generating global explanations of a variational autoencoder for anomaly detection.
Their approach applies a gradient-based fingerprinting technique that explains which features contribute to anomalies for various kinds of attacks.

Our proposed \sys system shares the same design goal with global-explanation approaches.
However, instead of relying on automated machine learning that is devoid of human judgement and domain knowledge, our visual analytics approach is designed to keep the analyst in control so that they can discover the \textit{global explanations} that are relevant and meaningful to their domain and analysis context.

\subsection{Visual Analytics for Anomaly Detection and Analysis} 
The visual analytics community has a long history in the research and development of systems for anomaly detection, dating back the original visual analytics research agendas proposed by Thomas and Cook\cite{thomas2006visual} and Keim et al.\cite{keim2008visual}.
However, most of these visual analytics systems have been designed to help analysts detect anomalies and outliers (e.g., in fraud detection\cite{leite2017eva, leite2018visual, chang2007wirevis} and network security monitoring\cite{mansmann2007visual, goodall2009viassist, xiao2006enhancing}), or inspect machine learning models to improve detection accuracy, reduce false positives, debug models, etc.\cite{wexler2019if, choo2018visual, hohman2018visual}.
Few systems exist that directly supports the analyst in discovering the potential reasons behind the anomalies and reporting the findings to business stakeholders and decision makers.

Some notable exceptions include the Situ system\cite{Goodall2018Situ}, which provides cyber network analysts with a contextual understanding of suspicious behaviors to aid in decision making.
Similar to our design, Situ also utilizes unsupervised anomaly detection algorithms for identifying potential suspicious data and provides analysts with an interactive interface to make sense of the anomalies.
Where our approach differs is that our system automatically generates potential explanations, thus reducing the analysts' need to synthesize the anomalies to gain the contextual understanding.

Other visual analytics systems exist for helping analysts explore and gain insight into the characteristics of data, which inevitably includes detecting and discovering outliers.
Examples include systems in the domains of space-time and trajectories analysis (e.g.\cite{andrienko2003exploratory, aigner2011visualization, liao2010anomaly, maciejewski2009visual}), network security (e.g.\cite{Xu2020CloudDet, Zong20193DNetworkIntrusion, Xie2019CST} and surveys by Shiravi et al.\cite{shiravi2011survey} and Zhang et al.\cite{zhang2017survey}), and systems such as Voila\cite{Cao2017Viola} -- used in urban development scenarios -- and TargetVue\cite{Cao2016TargetVue} -- used in social media analysis -- that can help an analyst discover suspicious behavior in dynamic and heterogeneous data by comparing current event sequences to statistical models of previously observed data. 

Similar to Situ, most of these examples are designed for specific domains. Further, they emphasize fluid data exploration but often lack support to assist an analyst in reasoning about anomalies and reporting the findings. 
\sys differs from the prior work in that it is designed to be domain-agnostic and focuses specifically on anomaly reasoning and reporting.
In the next section, we describe the challenges and design requirements for a visual analytics system designed for anomaly reasoning.

\section{Design Requirements and Motivation} 
\label{sec:motivation}
The design of \sys is inspired and informed by interviews with five professional data analysts at an (automotive) insurance company and one analyst in the health care industry.
All analysts have jobs that involve anomaly detection and reporting.
Each of the interviews lasted 30 minutes and was conducted over video conferencing due to COVID-19 restrictions.
During the interviews, participants were asked to describe their duties, problems that they frequently face, and tools or techniques that they commonly use for their tasks. 

\subsection{Analysts' Workflow and Practice} 
All six analysts describe a similar anomaly detection workflow. 
First, most (5 out of 6) analysts use off-the-shelf machine learning techniques to detect anomalies in their data (with one analyst using custom queries and scripts).
The analysts then manually inspect the resulting anomalous data points, sometimes with the use of visualizations and additional machine learning techniques (e.g., clustering).
All analysts reported exploring data in detail within Jupyter notebooks, or similar interactive programming environments. These environments allow a number of useful operations, including generating statistical summaries, filtering, and visualization. Most (4 out of 6) analysts also reported using commercial data exploration tools, like Tableau or Qlik, for investigating data.
Ultimately, all analysts report that their goal with anomaly detection is to discover patterns that are generalizable and can be reported to their managers or business stakeholders. 
In their experience, reporting individual anomalous data points is not useful in practice because the managers and stakeholders cannot take meaningful actions without some higher-level explanation of the anomalies.

\subsection{Gaps and Needs in the Workflow} 
Across all the steps in the analysts' workflow, the analysts report three tasks that are the least supported and therefore most burdensome.
First, the analysts describe the need to \textbf{organize} the anomalous data points (found by the anomaly detection algorithms) into meaningful groupings.
Analysts report that grouping similar anomalies together helps them to formulate hypotheses about the root causes of the anomalies.
However, this process is time-consuming because there can be nearly infinite ways to group and interpret the data.
The analysts need to explore these groupings through a series of trial-and-errors before finding the most meaningful groupings.

Second, the analysts discuss the need to manually \textbf{inspect} the outputs of the automated anomaly detection algorithms. 
The purpose of the inspection is two-fold: (1) the analysts note that anomaly detection algorithms lack domain knowledge. 
Sometimes flagging a data point as anomalous is accurate from an algorithmic standpoint, but would be inappropriate given the domain context (and vice versa).
Resolving such ambiguities requires human knowledge and domain expertise. 
(2) Through the inspection process, the analyst can build trust and confidence in the found anomalies. 
No analysts in our interviews felt comfortable reporting the outputs of an anomaly detection algorithm without inspecting the results first.
Analysts currently rely on manual programming in tools such as Jupyter notebooks to inspect the data.
Integrated support for comparing groups of anomalies that combines visualization with scripting will greatly reduce the analysts' effort when inspecting and building trust in their data.

Lastly, all analysts note the difficulty in producing \textbf{reports} for stakeholders and decision makers.
Since stakeholders and decision makers are often not savvy in data science, they require explanations at a higher semantic level that are not just statistical probabilities, distributions, and p-values.
Producing these high-level explanations is difficult because these reports need to be connected to business logic, the explanations need to be human-readable, and there needs to be intuitively understandable evidence to support the findings.
Currently there is no tool that can support the analysts in this report-preparation task and each analyst completes this step in their own \textit{ad hoc} fashion. 

\subsection{Design Requirements} 
\label{sec:requirements}
Our interviews with the analysts highlighted the current technology gaps in supporting their anomaly reasoning and reporting tasks.
Based on the feedback from the analysts, we summarize the currently unsupported tasks into three design requirements. 
We assert that a visual analytics system for anomaly reasoning must support these requirements:

\begin{enumerate}[topsep=6pt, partopsep=0pt,itemsep=3pt,parsep=0pt, label=\textbf{(R\arabic*)}]
\item \textbf{Organize}: The system must support an analyst in organizing anomaly data points found by a detection algorithm. Ideally the system can suggest multiple and diverse organizations that might suggest possible reasons behind the anomalies.
\item \textbf{Inspect, Refine, and Build Trust}: The system must support the analyst in building trust in the found anomalies by allowing the analyst to inspect and refine the anomalies. 
\item \textbf{Report}: The system must support the analyst in discovering higher-level explanations of the anomalies such that these findings can be reported to stakeholders and decision-makers who might not be savvy in data science.
\end{enumerate}

\section{\sys System Overview} 
\label{ref:pixal}

There are three components to \sys.
Each component addresses one of the three design requirements, which are: (R1) the \alg: an algorithm for automatically grouping anomalous data points into meaningful groupings, (R2) \sysvis: a visualization tool for supporting an analyst in inspecting and building trust in the found anomalies, and (R3) \sysreport: a visualization tool for helping an analyst generate hypotheses and a report of their findings.

Figure~\ref{fig:analystWorkflowAndRequirements} shows a high-level overview of how \sys supports an analyst's workflow.
Prior to using \sys, an analyst would run an anomaly detection algorithm over the input data, resulting in an \textit{anomaly score} for each of the data points.
\sys is agnostic to the specific algorithm used to generate the scores and can accept anomaly scores as either numeric values or binary labels.

The \alg takes the data points and the accompanying anomaly scores as input and generates \textit{groupings} of anomalies.
The groupings are represented using first-order predicate logic.
An important property of the \alg is that it can generate multiple and possibly overlapping predicates, %
which can be used by analysts to explore different plausible reasons behind the anomalies. 

Second, the \sysvis visualization takes the (multiple) predicates generated by the \alg and allows the analyst to inspect, explore, select, and refine them.
\sysvis provides visualizations and interaction techniques that support the analyst in performing counter-factual and comparative analyses to assess and build trust in the predicates.
Irrelevant or meaningless predicates can be discarded during this process while potentially interesting predicates can be fine-tuned and examined in more detail.

The output of \sysvis is a set of ``verified'' predicates, which the \sysreport visualization tool takes as input.
As these predicates only describe what data points are included in the predicate (i.e., the \textit{what}), the purpose of \sysreport is to help the analyst discover the possible reasons behind these anomalies (i.e, the \textit{why}).
In addition to helping the analyst test hypotheses, \sysreport can be used to create 
reports that are suitable for a stakeholder or a decision-maker who might not have expertise in data science.

In the following sections we describe each of the components in more detail.

\section{Recursive Predicate Induction Algorithm}
Conceptually the recursive predicate induction (RPI) algorithm is similar to that of a decision tree (for discrete labels) or a regression tree (for real values) in that it is meant to generate ``human readable'' groupings of (anomalous) data points.
In order to support anomaly reasoning, the \alg uses two interlocked processes to generate groupings.

\subsection{Generating Multiple Anomaly Groupings} 
\label{sec:algo:predicate}
Inspired by systems like
Scorpion\cite{wu2013Scorpion}, DIFF\cite{Abuzaid2018Diff}, iDiff\cite{Sarawagi2001iDiff}, Slice Finder\cite{Chung2020AutomatedDS}, and the Cascading Analysts Algorithm\cite{Ruhl2018CascadingAnalysts}, \alg uses first-order predicate logic to represent groupings of anomalous data.
The idea is to recursively partition the data using values along the data dimension as ``split points.''
However, instead of having one split point per decision level as decision trees or regression trees, the use of predicates allows for ``partitioning'' along the dimensions, resulting in first-order features such as (\texttt{10 < var < 20}) or \texttt{(id in [1, 3, 10])}. 
When recursively applied, the features produced by our algorithm form a predicate $p$: a union of $k$ first-order features:
\begin{align}
p = \cup_{i=1}^{k} \{f_i\}
\label{eq:predicate}    
\end{align}

The \alg follows a bottom-up recursive process similar to Slice Finder\cite{Chung2020AutomatedDS}. 
Specifically, the algorithm starts with initializing $n=\sum_{i=0}^{d} b_i$ ``\textit{base predicates}'', where $d$ is the number of data dimensions and $b_i$ is the cardinality of the data dimension $i$ if it contains categorical data and number of bins if it contains numeric values (note: bin size is a hyperparameter of the algorithm).
At each step of the recursion, a predicate $p$ is expanded to include another base predicate, resulting in a new predicate $p^{\prime}$. 
The Bayesian hypothesis score (see next Section, \ref{sec:algo:bayes}, for detail) is calculated for both $p$ and $p^{\prime}$.
If $p^{\prime}$ has a better score than $p$, $p$ is replaced with $p^{\prime}$ and the recursion continues.
Otherwise $p$ joins with another base predicate and is reevaluated.
If $p$ cannot be improved further, $p$ is returned.


For optimization, all $n$ base predicates are expanded in parallel.
After each step of the recursion, all predicates are considered for merging if two predicates are the same (e.g., $p_1 = \{f_a, f_b\}$ is the same as $p_2 = \{f_b, f_a\}$ since the features in a conjunctive first-order predicate are order invariant). 




\subsection{Bayesian Hypothesis Testing}
\label{sec:algo:bayes}
Given a predicate $p$, the purpose of Bayesian hypothesis testing is to determine if $p$ describes a subset of data that is statistically significantly more anomalous than the rest of the data. 
Conceptually, this can be described as a function $g$: 
\begin{align}
K = g(D, p(D))
\label{eq:p-subset}    
\end{align}
where $p(D)$ represents the subset of data instances that satisfy the predicate $p$ as it is applied over the dataset $D$. 
$K$ is the metric -- if it is greater than some threshold, $p(D)$ will be considered to be significantly more anomalous than $D$.
With Bayesian hypothesis testing, $K$ is formally defined as:
\begin{align}
K = \frac{Pr(\mathcal{D}|\mathcal{H}_{1})}{Pr(\mathcal{D}|\mathcal{H}_{0})}
\label{eq:hypotheses}
\end{align}
where $\mathcal{D}$ is a set consisting of both the full dataset $D$ and the subset of data $p(D)$ such that $\mathcal{D} = \{D, p(D)\}$. $H_{0}$ is the null hypothesis that the subset $p(D)$ is not different from the rest of the data $D$, and $H_1$ is the alternative hypothesis (that $p(D)$ is different from $D$). 


$K$, as expressed in Equation~\ref{eq:hypotheses}, is commonly referred to as the \textit{Bayes factor}\cite{Kass1995BayesFactor}.
Since $K$ represents the \textit{odds}, or the \textit{likelihood ratio} of the (marginal) likelihood of the two competing hypotheses, the values of Bayes factors are consistent in that they are invariant to differences in the definition of anomaly score.
As such, there is a universal interpretation of the values of $K$.
$K$ less than 3.2 suggests \textit{no} or \textit{bare} evidence for the alternative hypothesis ($H_1$).
Values between 3.2 and 10 suggests \textit{substantial} evidence; 10 to 100 is \textit{strong} evidence; and $K> 100$ is \textit{decisive} evidence\cite{jeffreys1998theory}. 
Compared to traditional hypothesis testing (such as p-values used in Slice Finder\cite{Chung2020AutomatedDS}), Bayes factors can express evidence to different degrees beyond just an accept/reject threshold.

In the \alg, we compute Equation~\ref{eq:hypotheses} using the \textit{JZS Bayes factor} introduced by Rouder et al.\cite{BayesTtest}, which has been implemented as a library available in R\footnote{\url{https://cran.r-project.org/web/packages/BayesFactor/index.html}}.
For the full definition of the \textit{JZS Bayes factor} as a Bayesian hypothesis test, we refer to Equation 1 in the work by Rouder et al.\cite{BayesTtest} (p. 231).

In the context of anomaly reasoning, we note a number of benefits with using the Bayes factor for determining the similarities between subsets of data.
First, higher values of $K$ represent more confidence in the explanation (that $D$ and $p(D)$ are different in their degrees of anomalousness). 
Second, the range of values of $K$ is consistent and invariant to the size or complexity of the data $D$. 
Lastly, $K$ increases not just as the average anomaly score in $p(D)$ increases but also as the number of data points in $p(D)$ increases.
This follows from the intuition that the more anomalous data points a grouping contains the more confidence we should have that the grouping is meaningful.

\subsection{Practicalities}
In addition to the benefits of Bayes factors, the other advantages of the \alg over other grouping algorithms is that the \alg provides the analyst with multiple overlapping groupings for the same data points.
However, being able to generate multiple overlapping groupings requires taking a recursive approach to searching the space of possible predicates rather than greedily maximizing Bayes factor.
While the \alg only takes approximately 10 seconds to run on small datasets (10k rows and 10 dimensions), larger datasets (100k rows, 30 dimensions) can take 20+ minutes, while large datasets (1 million rows, 100 dimensions) can take hours to complete.
The potentially long running time means that in practice the \alg is best suited for offline computation.
%
We discuss the use of the \alg in an interactive visual analytics environment further in the Discussion Section (Section~\ref{sec:disc:mixed}).



\section{\sysvis}
The output of the \alg is a set of predicates and their associated evidence scores ($K$).
With \sysvis, an analyst can examine these predicates in more detail.
The analyst can inspect each predicate, discard a predicate, (manually) generate new predicates, or refine (edit) an existing predicate.
These operations allow the analyst to perform counter-factual (i.e. ``\textit{what-if}'') analyses and comparisons between predicates to produce the final predicates for the reporting phase. 

\begin{figure}
\centering
\includegraphics[width=0.95\linewidth]{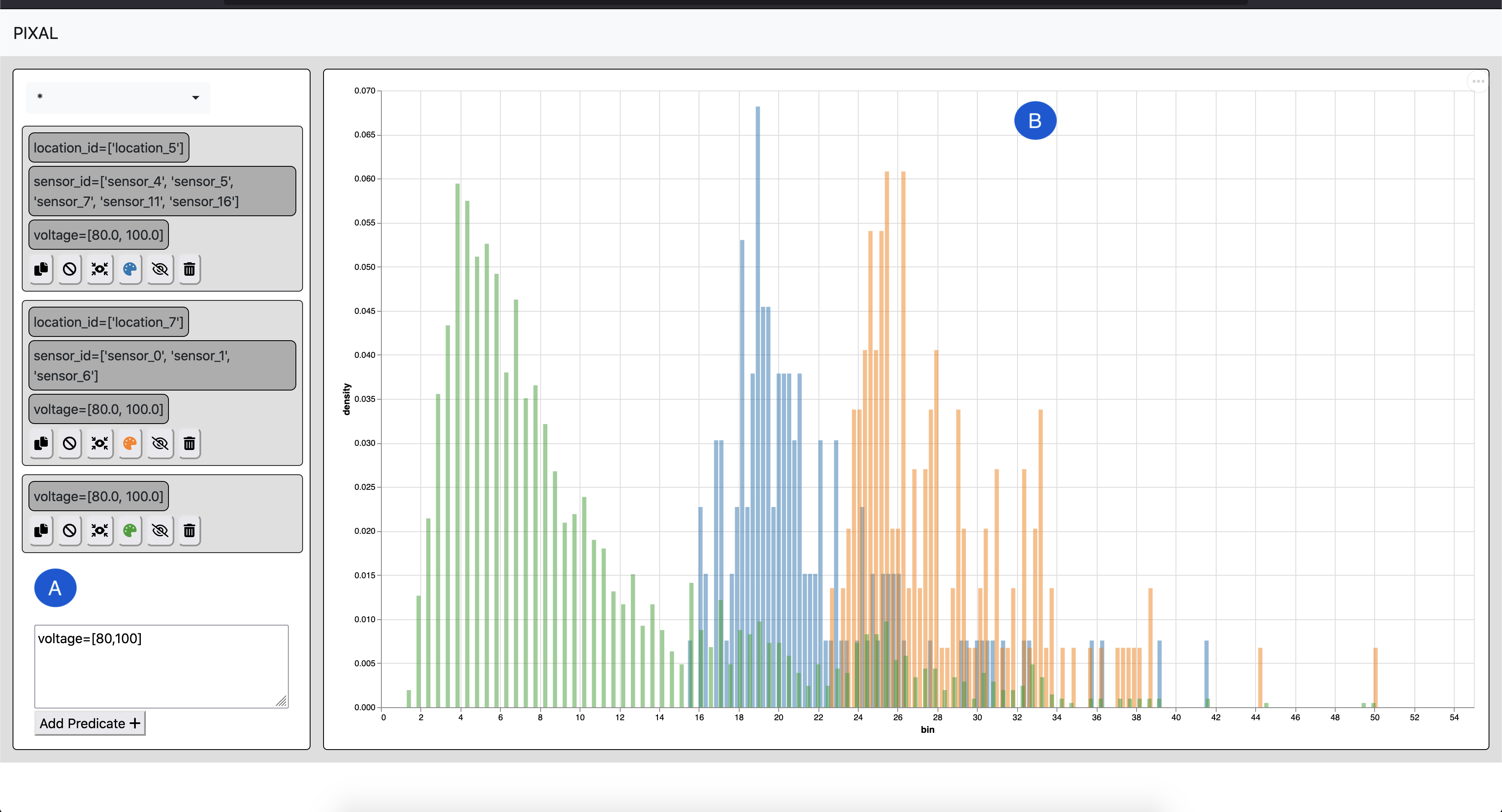}
\caption{\sysvis allows for: (A) the creation of new predicates and the modification of predicates generated by the \alg. (B) The distribution of anomaly scores in each predicate is rendered as a histograms. The x-axis of the histogram is the range of anomaly scores, and the y-axis is count. An analyst can compare predicates by super-positioning\cite{gleicher2017considerations} their corresponding distributions.
}
\label{fig:pixalexplorer_overview}
\end{figure}

As shown in Figure\ref{fig:pixalexplorer_overview}, \sysvis is a visualization tool that can render any first-order predicate (based on the input data) as a distribution in a histogram. 
The y-axis of the histogram represents count and the x-axis the range of the anomaly scores.
Multiple predicates can be visualized at the same time using different colors, allowing for superpositional comparisons of the predicates' distributions\cite{gleicher2017considerations}.

\vspace{3pt}
\noindent\textbf{Predicate Editing and Refinement:}
An analyst can manipulate a predicate in multiple ways.
First, a text box (shown in Figure~\ref{fig:pixalexplorer_editing}) allows the analyst to type in a predicate manually or select an existing predicate and edit its string representation.
Alternatively, the analyst can use the GUI elements supported by \sysvis to simplify the process. 
For example, a slider is provided for modifying the ranges of a numeric feature in a predicate, and a dropdown menu for adding or removing values of a categorical feature.
Not only do these GUI elements reduce the editing effort, they allow an analyst to quickly perform counter-factual analysis -- trying out different value ranges and observing the changes in the visualization interactively.


\vspace{3pt}
\noindent\textbf{Predicate Management:}
To aid the analyst in exploring and analyzing multiple predicates, \sysvis provides tools to help manage the (potentially many) predicates.
Figure~\ref{fig:pixalexplorer_editing} shows the panel where the analyst can hide a predicate, remove a predicate from the visualization, change the color of the predicate in the visualization, clone a predicate (to allow for experimentation), delete a predicate, and take the complement of a predicate (i.e., to select the data subset $(D - p(D))$ from Equation~\ref{eq:p-subset}).
%
After the analyst has completed their inspection and is satisfied with the final set of predicates, they can export the predicates for use in the next tool, \sysreport. 

\section{\sysreport}
The final step of the anomaly reasoning process is for an analyst to come up with hypotheses about the cause of anomalies and prepare a report for a stakeholder or decision maker.
Since stakeholders and decision makers are not always well versed in data science, the report needs to provide high-level explanations of the anomalies that the decision makers can relate to.

In order to support an analyst in generating such reports, \sysreport must support the analyst in two tasks.
First, the analyst must be able to identify the ``\textit{why}'' behind the anomalies that is meaningful to the stakeholder or decision maker.
Second, for each of the possible explanations, the analyst must be able to provide the evidence to support the finding.
Together, the ``\textit{why}'' and its accompanying evidence form a hypothesis about the potential reasoning behind the anomalies that a decision maker can act on.
During the reporting process, an analyst often include multiple plausible hypotheses in their reports.

Figure~\ref{fig:systemoverview} illustrates the visualization interface of \sysreport to support these two tasks.
It consists of five panels, which we describe below.

\vspace{3pt}
\noindent \textbf{Predicates View (Panel A)}: This view lists all the predicates produced by \sysvis. 
The predicates are sorted by their evidence scores. 
To begin the investigation and start forming hypotheses, the analyst selects a predicate in this panel, which will update all the subsequent views.

\vspace{3pt}
\noindent \textbf{Distribution View (Panel B)}: This view provides a visual confirmation of the distribution of the selected predicate. 
The Distribution View is the same as the histogram view in \sysvis except that it only shows two distributions. 
The blue distribution represents the selected predicate (i.e., $p(D)$ in Equation~\ref{eq:p-subset}) and the grey distribution its complement (i.e., $(D - p(D))$).


\vspace{3pt}
\noindent \textbf{Pivot View (Panel C)}. The purpose of the Pivot View is to support the analyst in investigating each of the \textit{features} in a predicate in more detail (see Equation~\ref{eq:predicate}).
For example, given a predicate \texttt{(locationId='location5') \& (sensorId in ['sensor4', 'sensor5', 'sensor7]) \& (voltage > 80)}, the analyst can \textit{pivot} on the \texttt{locationId}, the \texttt{sensorId}, or the \texttt{voltage} feature.
Pivoting on the \texttt{locationId} feature results in a barchart visualization where each bar represents a location id in the data (with the bar for 'location5' shown in blue, see Figure~\ref{fig:pixalate_pivot}). 
The y-axis of the barchart represents the average anomaly score.

Importantly, the height of each bar is only based on the subset of the data selected by the predicate.
For example, the bar representing the locationId 'location5' is based on the data selected with the predicate
\texttt{(sensorId in ['sensor4', 'sensor5', 'sensor7']) \& (voltage > 80)}
In this sense, the features not selected as the pivot become a \textit{filter} on the overall data. 

\vspace{3pt}
\noindent \textbf{Exploration View (Panel D)}: Once the analyst selects a pivot, the exploration view shows the detail of this selection and allows the analyst to explore further.
Inspired by visualization systems such as Tableau\cite{stolte2008polaris} and Polestar\cite{wongsuphasawat2015voyager}, with the Exploration View the analyst can select different visualization encoding (e.g., change the x or y-axes, the mark type, the colors, or the \textit{filters}). 

Initially the Exploration View defaults to using anomaly score as the y-axis -- the same as the visualization shown in the Pivot View.
As the analyst explores different attributes for the y-axis, they can identify possible reasons that make the predicate anomalous. 
For example, in Figure~\ref{fig:systemoverview}, the selected pivot in the Pivot View (Panel C) shows that the bars representing `sensor4', 'sensor5', and 'sensor7' (shown in blue) have significantly higher anomaly scores than the other sensors.
In the Exploration View (Panel D), the analyst has changed the y-axis to \texttt{temperature}.
From this visualization, the analyst observes that all three sensors have higher values in \texttt{temperature}, suggesting that the reason for the high anomaly scores for this predicate may be due to increased ambient temperature.

Lastly, the analyst can \textbf{bookmark}
a visualization.
The visualization can be exported later as the supporting evidence of the analyst's finding. 

\vspace{3pt}
\noindent \textbf{Recommendation View (Panel E)}: 
While the Exploration View supports the analyst in manual exploration, the Recommendation View automatically generates possible reasons for the anomalies in a predicate.
These recommendations are generated by computing the correlation between the anomaly scores of a predicate (given a pivot) and the attributes in the input data.
Attributes that correlate with anomaly scores (where the correlation coefficient is above $.3$) are added to the list of recommendations.


The recommendations are presented as both a natural language sentence and a visualization.
The natural language sentence is generated using a simple grammar that converts a first-order predicate into a sentence.
For example, the recommendation
``\textit{Average humidity is high when locationID is location5 compared to other locationID's when sensorID is sensor4, sensor5, and sensor7 and voltage is greater than 80}'' corresponds to the predicate: \texttt{(locationID = location5) \& (sensorID in ['sensor4', 'sensor5', 'sensor7']) \& (voltage > 80)}
with the pivot on the {LocationID feature and the attribute Temperature as the possible reason recommended by \sysreport for the anomalies in the predicate 

When an analyst clicks on a recommendation, the visualization will be displayed in the Exploration View for further examination and can be bookmarked for later exportation into a report.

\begin{figure}[t]
\centering
\begin{subfigure}[b]{0.48\linewidth}
\centering
    \includegraphics[width=\linewidth]{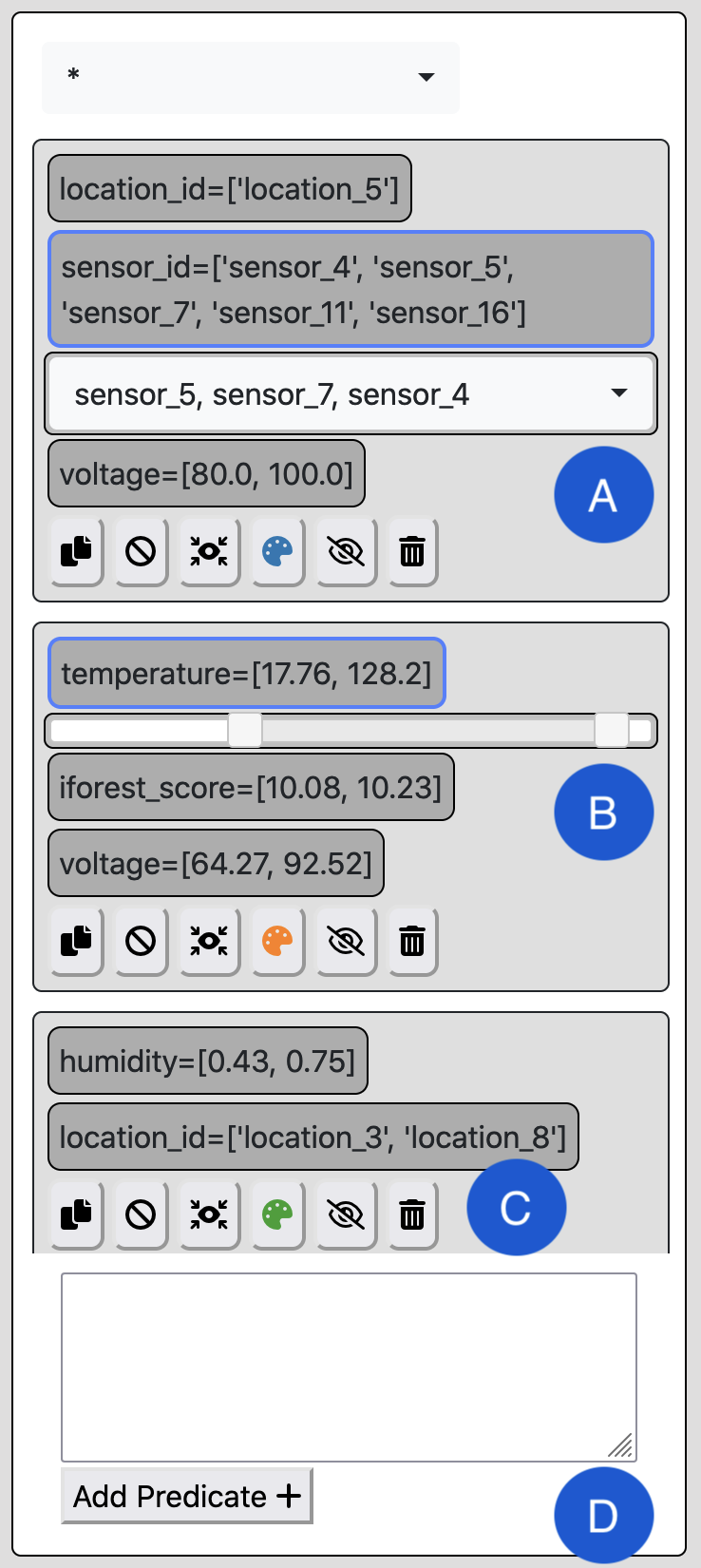}
    \caption{\sysvis Predicate Panel}
    \label{fig:pixalexplorer_editing}
\end{subfigure}%
~~
\begin{subfigure}[b]{0.48\linewidth}
\centering
\includegraphics[width=\linewidth]{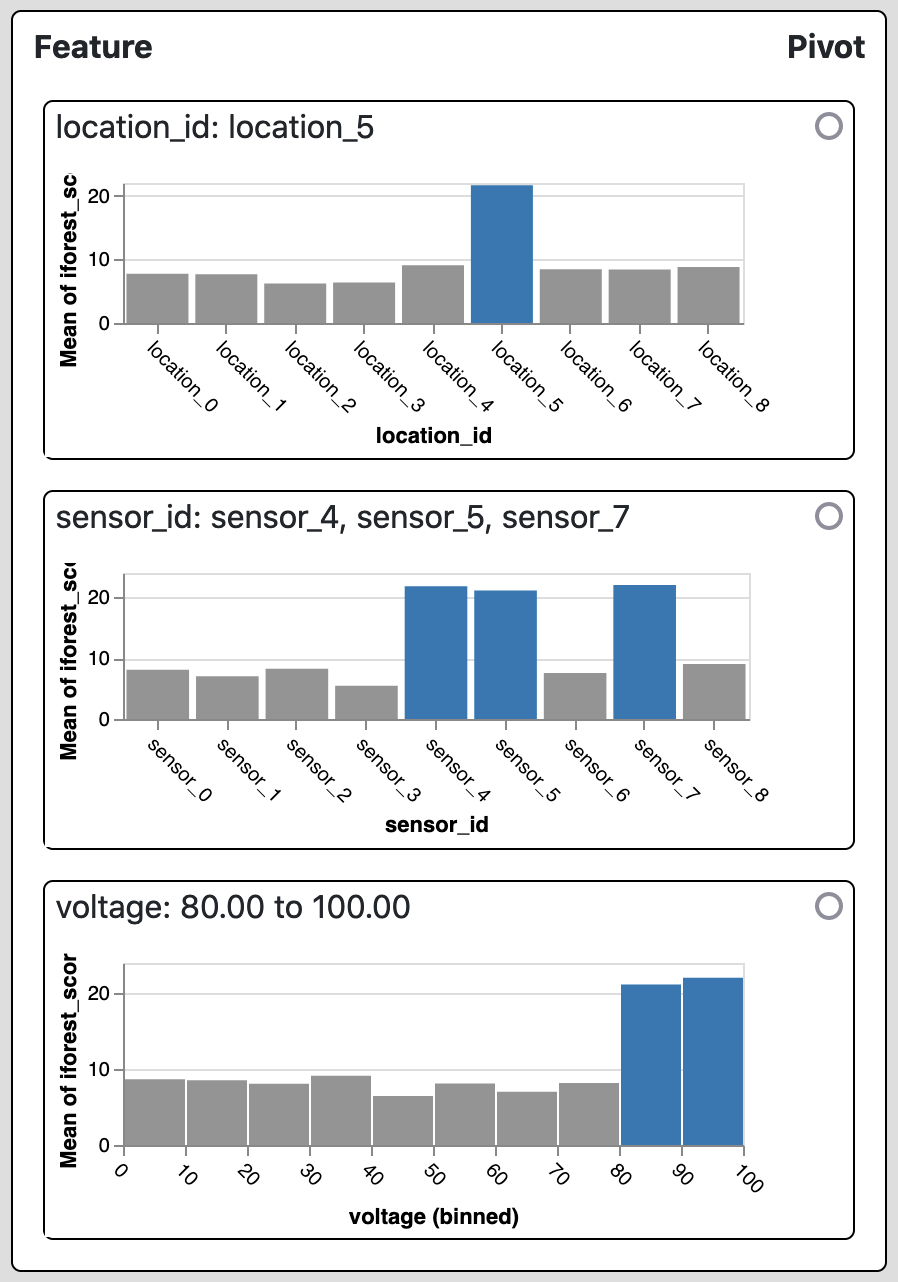}
    \caption{\sysreport Pivot View}
    \label{fig:pixalate_pivot}
\end{subfigure}
\caption{(a) \sysvis allows the user to edit and refine predicates with GUI elements ((A) a dropdown menu for choosing categorical values or (B) a slider for choosing the range of a numeric feature). (C) The user can also use this panel to manage the predicates, including abilities to hide, remove, duplicate, or take a predicate's complement. (D) Additionally, the user can add a new predicate or edit a one by manually typing into a text box. (b) The Pivot View. An analyst can examine anomaly scores from different perspectives by pivoting on any feature included in the selected predicate}
\label{fig:manmade}
\end{figure}

\section{Usage Scenario}\label{sec:usagescenario}
We describe our tool with a usage scenario, where an analyst uses \sys to reason about anomalies in a sales dataset.
The dataset used in this scenario is modified from the ``\textit{Superstore Sales}'' dataset included with Tableau Desktop (version 2020.4).
Each row in the dataset includes the \texttt{Sub-Category} of product sold (e.g. Furniture, Office Supplies), the \texttt{State} it was sold in, the customer \texttt{Segment} (e.g. Consumer, Home Office), \texttt{Order-Date}, the \texttt{Quantity} of the product being sold, and the \texttt{Unit-Price} and \texttt{Unit-Cost} of the product being sold.
Additionally, the data includes the average \texttt{Temperature} and \texttt{Precipitation} on the sales date.
In this scenario, an analyst has been tasked by Superstore executives with identifying and reporting anomalous phenomenon in the sales dataset.

The analysts begins by generating anomaly scores for the dataset.
The popular open source Python library Scikit-learn\cite{sklearn} includes a number of anomaly detection algorithms, including the isolation forest (iforest) algorithm\cite{liu2008isolation}.
Understanding that the Superstore executives will be primarily interested in sales related anomalies (i.e., \texttt{Quantity}, \texttt{Unit-Price}, and \texttt{Unit-Cost}), the analyst runs the iforest algorithm on these features, which then assigns each row in the data an anomaly score.

Next, the analyst imports the Superstore data and the assigned anomaly scores into the \alg, which produces 7 predicates (ranked by their evidence scores from high to low):

\smallbreak 
\begin{lstlisting}
 1. Sub-Category in ['Tables']
 2. Sub-Category in ['Machines']
 3. Sub-Category in ['Copiers']
 4. State in ['New York', 'Massachusetts', 'New Hampshire'] 
    & Segment in ['Consumer', 'Home Office'] & 
    2016-11-8 < Order-Date < 2017-1-21
 5. 2.74 < precipitation < 4.59 & -0.09 < Temperature < 
    36.36 & Segment in ['Consumer', 'Home Office']
 6. 2016-11-8 < Order-Date < 2017-1-21 & State = 'Vermont' 
 7. State = 'Vermont' & Segment = 'Corporate'   
\end{lstlisting}   
\smallbreak 


\subsection{Discovering Business Logic}
The analyst notices that the top three ranked predicates are all related to product \texttt{Sub-Category}.
This is consistent with the analyst's understanding that the Sub-Category of the product being sold will have a significant effect on the \texttt{Quantity}, \texttt{Unit-Price}, and \texttt{Unit-Cost}.
The analyst visualizes the three predicates in \sysvis and notices that the predicates have similar distributions.
Given the similarities between them, the analyst decides to visualize the three predicates together.
Using the text editor in \sysvis, the analyst manually types in a new predicate by combining the three, resulting in: 
\smallbreak 
\begin{lstlisting}
 8. Sub-Category in ['Tables', 'Machines', 'Copiers']
\end{lstlisting}   
\smallbreak 
The analyst can now see the complete distribution of anomaly scores for the three \texttt{Sub-Categories} together.
To verify that this predicate is indeed anomalous, the analyst compares this predicate against the rest of the data.
The analyst completes this goal by first cloning predicate 8, and clicking on the ``NOT'' button in the panel, resulting in a new predicate:
\smallbreak 
\begin{lstlisting}
 9. NOT(Sub-Category in ['Tables', 'Machines', 'Copiers'])
\end{lstlisting}   
\smallbreak 
Figure~\ref{fig:pixalexplorer_three_subcategories} shows the analyst's superposition of predicates 8 and 9 to compare their differences. 
From this visualization, the analyst can see that Machines, Copiers, and Tables (shown in orange) tend to have higher anomaly scores when compared to all other Sub-Categories (shown in blue).
\begin{figure}[t]
\centering
\includegraphics[width=0.95\linewidth]{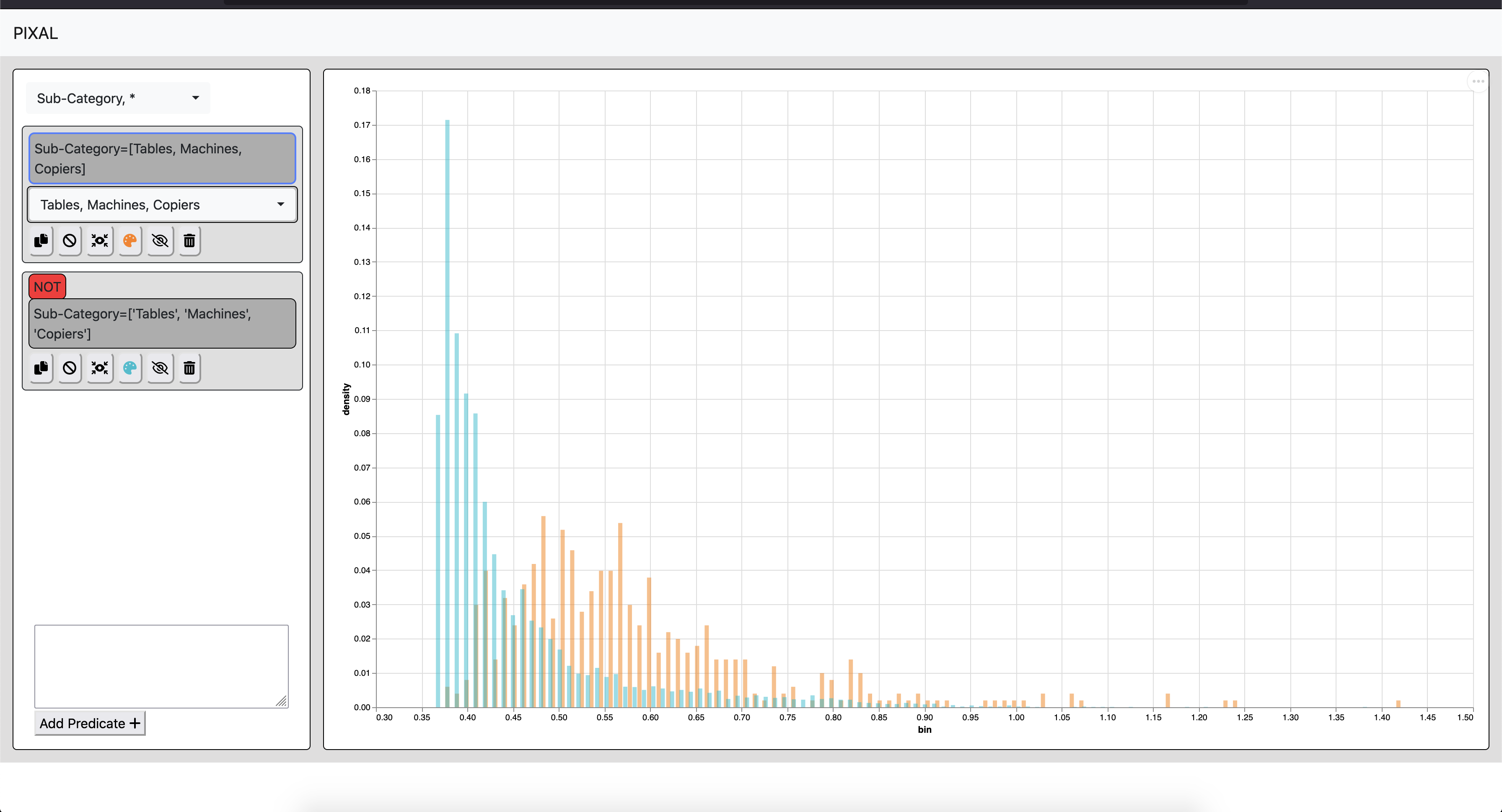}
\caption{An analyst compares predicate 8 (orange) and its complement predicate 9 (blue). Observing that anomaly scores are generally higher in predicate 8, they are able to confirm that this predicate contains anomalies and is worth further investigation.}
\label{fig:pixalexplorer_three_subcategories}
\end{figure}

To understand the potential reason behind this anomaly, the analyst copies predicate 8 and opens it in \sysreport. 
First, the analyst confirms with the Distribution View that anomaly scores are high for Tables, Machines, and Copiers compared to other Sub-Categories.
Turning to the Recommendation View, the analyst sees from the visualization and the text description that \texttt{Unit-Cost} and \texttt{Unit-Price} are both higher for Copiers, Machines, and Tables compared to other Sub-Categories (Figure~\ref{fig:pixalate_1b}).


While the analyst has confirmed that the sales of Copiers, Machines, and Tables are indeed ``\textit{data outliers}'' when viewed from the business operation standpoint, the fact that Copiers, Machines, and Tables have higher production costs (\texttt{Unit-Cost}) and therefore higher prices (\texttt{Unit-Price}) is not surprising.
The analyst still makes a bookmark of this finding in \sysreport, but decides not to report it to the decision makers because this finding merely reflects standard business logic.
Satisfied with this conclusion, the analyst returns to \sysvis to explore other predicates.
\subsection{Discovering Unexpected Outliers}
Continuing the investigation, the analyst examines predicate 4 in \sysvis.
The analyst noticed that in addition to predicate 4, predicates 5, and 7 all contain the feature \texttt{Segment}. 
Visualizing predicate 4 in \sysvis, the analyst wants to examine and isolate the effect of \texttt{Segment}.

The analyst types in the text box in \sysvis and creates a new predicate that contains only the \texttt{Segment} feature from predicate 4, resulting in:
\smallbreak 
\begin{lstlisting}
  10. Segment in ['Consumer', 'Home Office']
\end{lstlisting}   
\smallbreak 

From this view, the analyst observes that predicate 10's distribution skews towards smaller anomaly scores and does not significantly overlap with predicate 4 and reasons that \texttt{Segment} itself is not the cause of the anomaly.
The analyst subsequently decides that the other two features in predicate 4 (\texttt{State} and \texttt{Order-Date}) will need to be examined closely.

Turning to \sysreport, the analyst first examines the \texttt{State} and then the \texttt{Order-Date} features of predicate 4 in the Pivot View.
When pivoting on \texttt{State}, the analyst observes a recommendation in the Recommendation View (see Figure~\ref{fig:pixalate_2a}) that suggests that the sales \texttt{Quantity} in New York, Massachusetts, and New Hampshire are lower compared to other states (between the \texttt{Order-Dates} 2016-11-8 and 2017-1-21, and in the Consumer and Home Office \texttt{Segments}). 


When pivoting on \texttt{Order-Date}, the analyst observes two recommendations that suggest that \texttt{Precipitation} values are higher and the \texttt{Temperature} values are lower during the period of interest.
Together with the previous finding, the analyst begins to form the hypothesis that the decreased sales \texttt{Quantity} in New York, Massachusetts, and New Hampshire during the dates between November and January are likely due to winter storms with high \texttt{Precipitation} and low \texttt{Temperature}.

Interestingly, looking ahead at predicate 5, the analyst sees the confirmation that high \texttt{Precipitation} and low \texttt{Temperature} of the same \texttt{Segments} is also found to be anomalous.
This observation gives the analyst increased confidence in their finding.
The analyst bookmarks these visualizations in \sysreport and includes them in their report.



\begin{figure}
\centering
\includegraphics[width=0.95\linewidth]{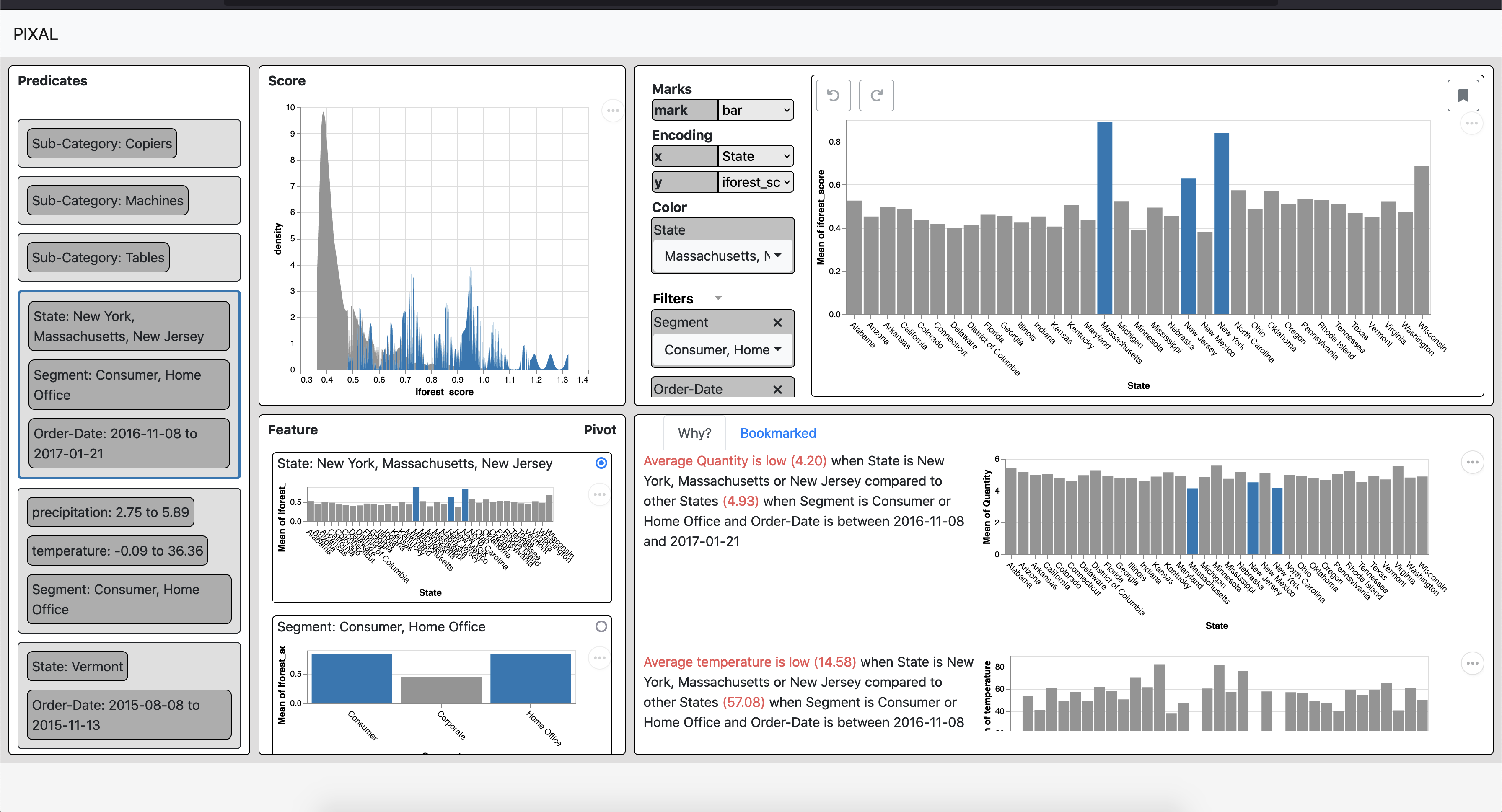}
\caption{This figure shows an analysis of a predicate with multiple features using \sysreport. Examining the feature relating to \texttt{State} as the pivot, an analyst identifies the increased \texttt{Quantity} in sales as a potential cause of anomalies in the Recommendation View.}
\label{fig:pixalate_2a}
\end{figure}

\section{Interview Study}
\label{sec:eval:observational}
We performed an interview study to evaluate the effectiveness of \sys in helping data analysts solve anomaly reasoning tasks.

\vspace{4pt}
\noindent\textbf{Participants}: 
We recruited three professional data analysts to participate in this interview study.
All three participants had experience with anomaly detection as part of their job duties.
Two of the analysts work for health care companies, with one of the two having also participated in the initial set of interviews (see Section \ref{sec:requirements}).
The third analyst works as a data scientist for an academic institution.



\vspace{4pt}
\noindent\textbf{Procedure}:
We walked each participant through a scenario where an analyst uses \sys to reason about anomalies in a sales dataset and reports their findings to decision makers.
After introducing the dataset, we demonstrated the use of \sys on the scenario described in Section~\ref{sec:usagescenario}.
The participants were encouraged to ask questions and direct the exploration process in the style of pair-analytics\cite{arias2011joint}.

Following the tasks each participant was interviewed to provide feedback on the usability of the tool, comparison to traditional anomaly detection workflows, and how well the explanations generated using \sys could be used to communicate to decision makers.
Each of these studies lasted 60 minutes and was conducted over video conference due to COVID-19 considerations.
With the consent of participants, videos of the interviews were recorded.

\begin{figure}
\centering
\includegraphics[width=0.95\linewidth]{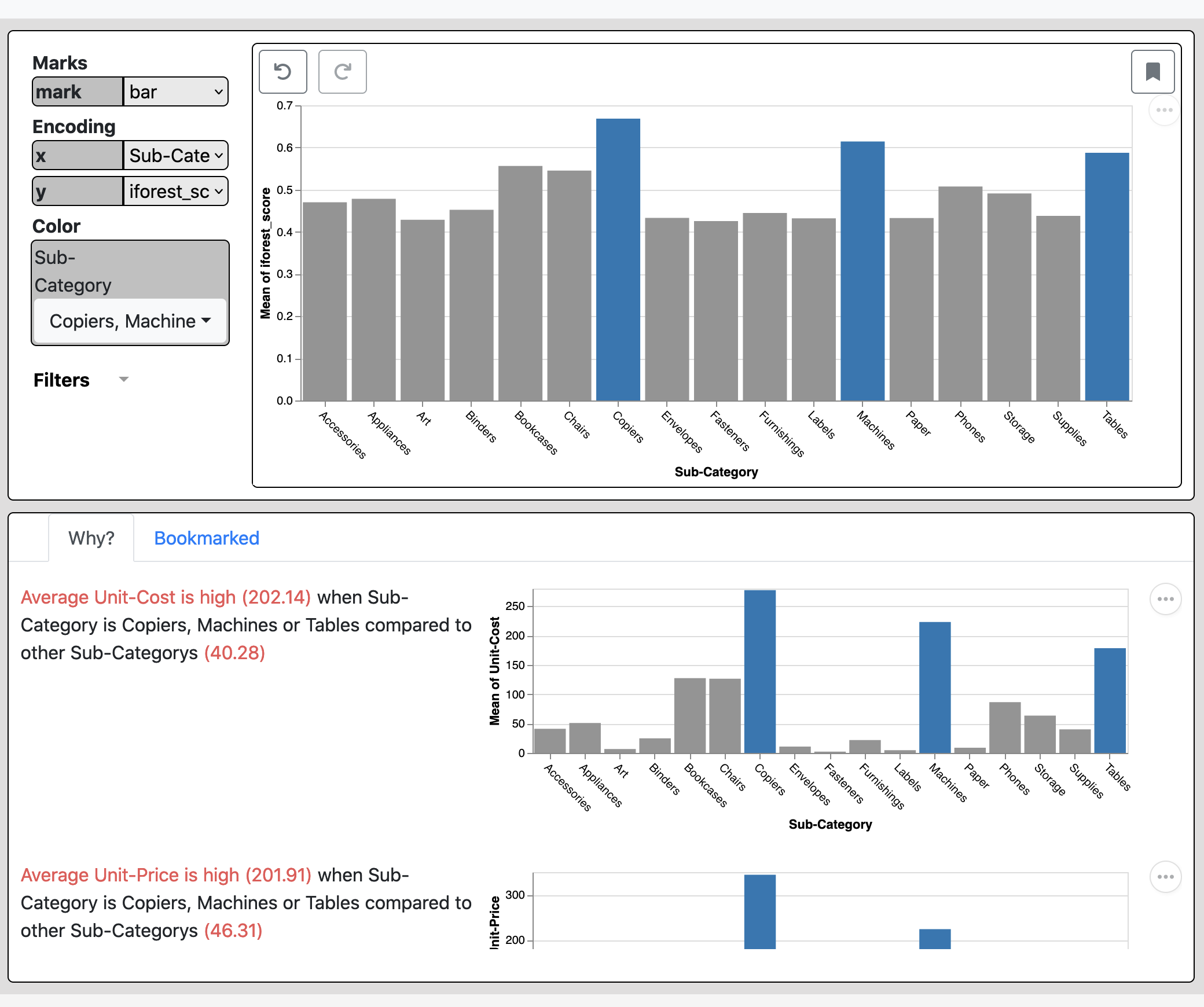}
\caption{The analyst examines the possible reasons for the anomalies relating to sales of products in the Tables, Machines, and Copiers \texttt{Sub-Category}. In the recommendation view (bottom), the \sysreport tool suggests two possible reasons behind the anomalies -- increased \texttt{Unit Cost} and increased \texttt{Unit Price}.}
\label{fig:pixalate_1b}
\end{figure}

\subsection{Study Result}
We reviewed the qualitative feedback from the interviews and distilled them into six main findings.
These findings are grouped by how well \sys supports the three design requirements (Section~\ref{sec:requirements}).

\vspace{3pt}
\noindent\textbf{R1 (Organize): Generating Anomaly Groupings:}
Two out of the three participants agreed that analyzing groups of anomalies is usually preferred over analyzing single anomalies:
``\textit{Individual anomalies happen frequently and do not necessarily corresponds to a persistent issue in the system}.''
``\textit{Someone who is really interested in the data might still investigate individual data points, but predicates will be sufficient for understanding the high-level context in most cases}.''

\vspace{3pt}
\noindent\textbf{R1 (Organize): Benefits of Multiple Predicates:}
In addition to bringing focus to persistent groups of anomalies, two out of three participants noted that predicates also have the advantage of being easy to understand:
``\textit{Understanding the data as predicates instead of the raw data makes it a lot easier to consume, even for someone who knows what they’re looking for}.''
``\textit{The predicates seem flexible and easy to understand. You’re really just talking about grouping with different features}.''

All three participants agreed that generating multiple predicates helped with anomaly reasoning:
``\textit{When we’re trying to explain a group of anomalies we test a collection of hypotheses, having multiple predicates lets you compare multiple different classes of anomaly types}.''
``\textit{Anomalies often occur as part of a convoluted chain of events. Multiple explanations are needed in these cases to get the full story}.''
``\textit{Having the ability to cut across the data in many different ways is always useful}.''

\vspace{3pt}
\noindent\textbf{R2 (Inspect, Refine and Build Trust): Building Trust: }
All three participants agreed that the tools provided by \sys would help build trust in the predicates:
``\textit{The more interactions a user has access to the more they will trust the system. \sys seems like it provides a complete set of tools for interacting with and building trust in predicates}.''
``\textit{The predicates generated by PIXAL seem easy to verify. I trust them more knowing that I can easily check whether those data points are actually anomalous or not.}''

\vspace{3pt}
\noindent\textbf{R2 (Inspect, Refine and Build Trust): Predicate Refinement: }
Participants described scenarios where predicates could be manually refined to more closely align with an application domain.
``\textit{Analysts need to be able to play with their data and experiment with different hypotheses that might not necessarily be presented by the system. These tools seem like they would help investigate alternatives}.''
``\textit{For example, the date ranges in the predicates generated by PIXAL do not necessarily align with time periods that are meaningful for the business. In this case it would be helpful to modify the predicate so that date ranges align with financial quarters or some other meaningful period}.''

Participants also found that being able to generate new predicates for comparison is useful in building trust:
``\textit{The only way we can establish anomalousness is comparison with known inliers. In our domain we have access to a lot of normal examples that could be used for comparison}.''
``\textit{A lot of times people already have an idea of what they’re expecting to see. The only way they’ll trust a system that’s telling them something that goes against their intuition is to confirm themselves}.''

\vspace{3pt}
\noindent\textbf{R3 (Report): Hypothesis Generation:}
Two of the three participants discussed the need to report hypotheses about the anomalies rather than individual anomaly points.
In particular, one participant in the healthcare domain said it was not feasible to report results directly from an anomaly detection algorithm:
``\textit{At this moment there is no accepted mechanism from a regulatory perspective enabling a system to determine what the anomaly is}.''

Participants further stressed the need for tools allowing human analysts to reason about hypotheses:
``\textit{We can determine from the output of an algorithm whether there is a tissue lesion or malformation in an image, but determining what exactly makes it abnormal must be done by a human}.''
``\textit{From a business perspective, when you see an anomaly, the first two things that will naturally come to mind are: is this going to happen again? And if so what can I do to prevent it? So being able to present hypotheses is valuable from that perspective}.''

Additionally, participants found that the tools provided by \sys helped them generate hypotheses in our case study:
``\textit{\sys highlighted three states as anomalous, New York, Massachusetts, and New Jersey. Knowing that anomalies also frequently occur when there is high precipitation and low temperatures, it’s likely that weather in those states in that time period caused a decrease in sales. If we can anticipate this type of weather in the future we could propose a $10\%$ off sale either for specific products or for online orders}.''
``\textit{This tool made it easy to reason about how a snowstorm on the upper east coast could be causing anomalies by interfering with sales}.''

\vspace{3pt}
\noindent\textbf{R3 (Report): Recommendations:} 
All three participants found the recommendations to be useful:
``\textit{The recommended text and plots work well together. You can quickly compare anomalies to normal data by reading the text and get a better sense of magnitude and overall pattern by looking at the plot}.''
In particular, participants found the text generated by \sys useful for explaining anomalies:
``\textit{An analyst might not be satisfied by a statistically fragile summary like average quantity, but usually this would be sufficient for reporting purposes}.''
They also found the visualizations generated by \sys useful for explaining anomalies:
``\textit{The plots help you quickly identify patterns and understand which features might be contributing to anomalies}.''

\vspace{3pt}
\noindent\textbf{Summary Feedback:}
All three participants stated that \sys would help them reason about anomalies in their job function:
``\textit{This seems like it would free up a lot of time usually spent digging into individual data points. Even if you are not looking for any particular anomalies this seems like a wonderful tool for exploratory data analysis}.''
``\textit{Pathologists need to be able to report what exactly is occurring that caused a section of an image to be flagged as abnormal. This system could help quickly identify patterns and understand which features might be contributing to anomalies}.''

\vspace{3pt}
\noindent\textbf{Suggestions for Improvement:}
Overall, the participants were enthusiastic about the design of \sys.
When asked about whether there are opportunities for improvement, one suggested a minor feature in the ability to export multiple bookmarks in the same report.
Another participant suggested that additional features in \sys could help an analyst better identify useful patterns in the anomalies:
``\textit{The plots generated by PIXAL could include additional features to highlight patterns in anomalies. For example, the reordering the x-axis so that anomalous categories are grouped together could help highlight differences between them and the normal data}.''

\section{Discussions and Design Implications}
The overall evaluation of \sys was very positive. 
The participants' feedback suggests that this design of \sys can effectively support analysts in performing anomaly reasoning. 
As \sys is developed following an iterative design process, there were prototypes that were not met with similar enthusiasm by during evaluation.
In this section we describe two of the lessons learned and the design principles that we distilled from the experience.

\subsection{Anomaly Reasoning as a Sensemaking Process}
From our interviews with the analysts and our observations of the participants interactions with \sys, we posit that the analysts' process of anomaly reasoning resembles the Sensemaking model proposed by Pirolli and Card\cite{pirolli2005sensemaking}.
Similar to Sensemaking, anomaly reasoning starts with a large amount of disorganized data.
Through the \textit{Foraging Loop} involving inspection, exploration, refinement, the analyst develops groupings of the anomalies. 
The analyst then undergoes the \textit{Sensemaking Loop} in which they develop mental models and hypotheses about the observed anomalies.
The process completes when the analyst produces a report (i.e. presentation) that \textit{tells a story} about the anomalies to stakeholders and decision makers.

The outputs from each component of \sys (see Figure~\ref{fig:analystWorkflowAndRequirements}) also correlate to the amount of \textit{Structure}
in the Sensemaking model. 
The initial inputs to \sys (i.e., the set of data points and their anomaly scores) lack both semantic meaning and structure. 
The \alg provides initial organization by grouping data into predicates.
The use of the \sysvis further allows the analyst to refine and select the most relevant predicates, shaping the predicates using their domain knowledge.
Finally, with \sysreport, the analyst summarizes the anomalies into succinct and human-readable sentences and visualizations that can be put into a report.

The similarity between the anomaly reasoning and the Sensemaking processes suggests a potential commonality in the design process.
We leave the confirmation and validation
of this observation for future work.


\subsection{Mixed-Initiative Anomaly Reasoning Considered Harmful}
\label{sec:disc:mixed}
During our iterative design process we
designed a prototype mixed-initiative visual analytics system for anomaly reasoning.
However, our evaluation with professional data analysts suggested that this approach could be potentially harmful.

In this prototypes, analysts were able to \textit{steer} the predicate generation process in the \alg. 
The analyst was able to suggest potentially interesting features to be included in the predicate (in real time), and the \alg would react by favoring those features when constructing the predicates.
Tn analyst was also able to terminate a predicate during the search process if the predicate is no longer of interest.
Figure~\ref{fig:old_design} shows the interface of this mixed-initiative prototype system.
\begin{figure}
\centering
\includegraphics[width=0.95\linewidth]{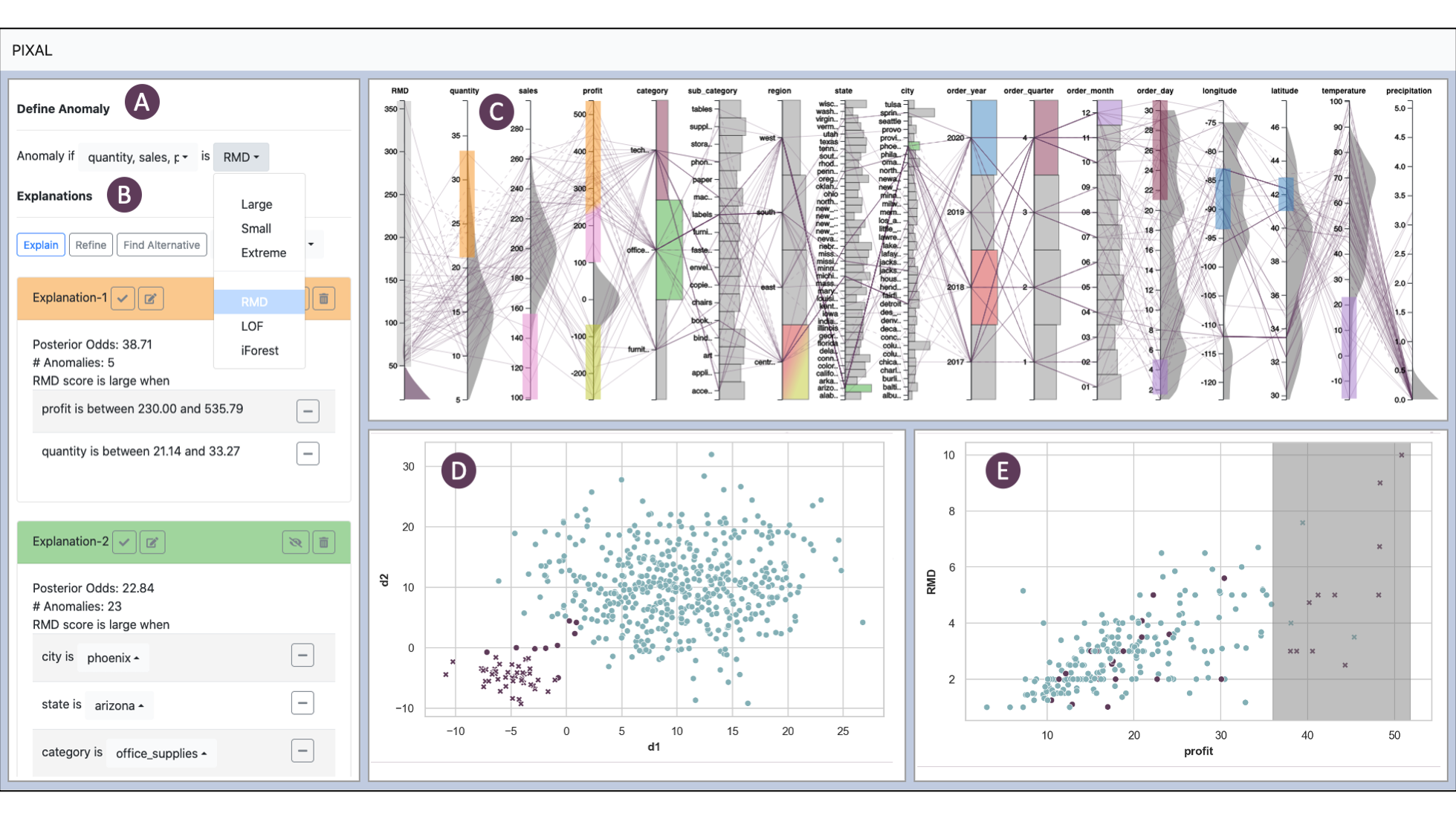}
\caption{A screen shot of a previous prototype of \sys. This prototype follows a mixed-initiative approach for an analyst to generate predicates. The left panel provides an analyst with a variety of ways to ``\textit{steer}'' the computation, which analysts found difficult to use. 
}
\label{fig:old_design}
\end{figure}

On the positive side, during an evaluation the participants noted that this prototype met the three design requirements as outlined in Section~\ref{sec:requirements}.
It addition, this approach had the benefit of much faster computation times from the \alg:
rather than searching an exhaustive space of predicates including potentially irrelevant features, the mixed-initiative approach allowed the analyst to prune irrelevant search spaces early. 

However, during the evaluation we also observed significant limitations to the mixed-initiative approach.
First, we found that the the participants were often unable to or were uncertain about \textit{how} to steer the predicates. 
For example, the participants were not able to decide with certainty whether a feature such as \texttt{precipitation} should be included in a predicate that is meant to detect anomalies about business cost.
Forcing the participants to make such decisions in real-time sometimes led them into making rushed and incorrect choices.

Additionally, analysts steering the computation would occasionally exhibit biases when generating the predicates.
If an analyst had a perceived assumption about what was likely causing the anomalies, their steering would often reflect the bias and direct the computation towards confirming that assumption.
Recent research has begun to explore the potential pitfalls of progressive visualization and steering (e.g.,\cite{procopio2021impact, cho2017anchoring, badam2017steering, wall2017warning, stolper2014progressive}).
In the context of anomaly reasoning, we observe that such bias can be detrimental to the analysts' abilities to appropriately reason about the anomalies.




\section{Limitations and Future Work}\label{sec:limitations}
Although the \sys benefited from the lessons learned from the iterative design and was found to be effective by professional data analysts, we note that there is still room for improvement. In this section we outline some of the limitations of \sys.

\vspace{3pt}
\noindent \textbf{Integrated Anomaly Detection:} 
One of the analyst expressed the desire to have integrated anomaly detection capabilities in \sys.
\sys currently assumes that the input data and their corresponding anomaly scores have been previously verified.
However, while the verification process can be performed as a preprocessing step, the analyst suggested that an integrated solution will allow for opportunities to iterate between anomaly detection and anomaly reasoning in cases where the anomaly detection algorithm performs poorly on the input data.

\vspace{3pt}
\noindent \textbf{Improved Recommendations:} During the evaluation the participants applauded \sys's ability to provide automated recommendations that explained the possible reasons behind the anomalies. 
We recognize that there are still opportunities to develop more sophisticated algorithms for improved recommendations.
To start, the recommendations in \sys are currently based on the correlation coefficients between the anomaly scores and each of the data attributes independently.
This limits \sys to only be able to identify single data attributes as potential reasons and excludes the interactions between multiple attributes as possible explanations. 
We will investigate other approaches, such as multi-linear regression and KL divergence in the future to improve the recommendations.

\vspace{3pt}
\noindent \textbf{Improved Natural Language Explanations and Visualizations:} 
As the recommendation algorithm becomes more sophisticated, the techniques for generating natural language explanations will need to be more complex.
Similarly, the ways that visualizations can be automatically generated to provide supporting evidence will also need to improve.
Explainable AI (XAI) is a burgeoning field in the visual analytics community. 
We will look to incorporate techniques developed in XAI research to further improve the visualizations in \sys in future work.

\section{Conclusions}\label{sec:conclusions}
In this paper, we present \sys, a visual analytics system for helping analysts reason about anomalies.
\sys fills gaps in existing tools for anomaly reasoning:
The \alg helps analysts find logical groupings of anomalous data points by generating and testing predicates using a Bayesian hypothesis testing framework.
\sysvis allows analysts to visualize and build trust in predicates generated by the \alg.
\sysreport helps the analyst generate and validate hypotheses that might explain the cause of a group of anomalies, and provides support for generating reports that can be presented to stakeholders.
We evaluated the utility of \sys with three professional data analysts who observed \sys being used to solve an anomaly reasoning problem.
All three analysts found that \sys facilitated anomaly reasoning by filling gaps in their current workflow.

\bibliographystyle{abbrv}

 \bibliography{main}
\end{document}